\documentclass{elsart}

\usepackage{cite}
\usepackage{graphicx}

\begin{document}

\begin{frontmatter}

\title{On nonadiabatic SCF calculations of molecular properties}

\author{Francisco M. Fern\'{a}ndez \thanksref{FMF}}

\address{INIFTA (UNLP,CCT La Plata-CONICET), Divisi\'{o}n Qu\'{i}mica Te\'{o}rica,\\
Diag. 113 y 64 (S/N), Sucursal 4, Casilla de Correo 16,\\
1900 La Plata, Argentina}

\thanks[FMF]{e--mail: fernande@quimica.unlp.edu.ar}

\begin{abstract}
We argue that the dynamic extended molecular orbital (DEMO) method may be
less accurate than expected because the motion of the center of mass was not
properly removed prior to the SCF calculation. Under such conditions the virial
theorem is a misleading indication of the accuracy of the wavefunction.
\end{abstract}

\end{frontmatter}

The first step in any quantum--mechanical treatment of atomic and molecular
systems is the separation of the motion of the center of mass. The
nonrelativistic Hamiltonian operator with only Coulomb interactions between
the constituent particles for such systems is of the form $\hat{H}_{T}=\hat{T%
}+V$, where $\hat{T}$ is the total kinetic--energy operator and $V$ is the sum of
all the Coulomb interactions between the charged particles. By means of a
straightforward linear combination of variables one rewrites the
kinetic--energy operator as $\hat{T}=\hat{T}_{CM}+\hat{T}_{rel}$, where $%
\hat{T}_{CM}$ and $\hat{T}_{rel}$ are the operators for the kinetic energies
of the center of mass and relative motion, respectively. Then one solves the
Schr\"{o}dinger equation for the internal Hamiltonian $\hat{H}=\hat{T}%
_{rel}+V$\cite{BD03,KW66,KA00}.

It is well known that the eigenfunctions of $\hat{H}_{T}$ are not square
integrable. For this reason, it is at first sight striking that Tachikawa et
al\cite{TMNI98,TO00} carried out their dynamic extended molecular orbital
(DEMO) method on the total Hamiltonian operator $%
\hat{H}_{T}$. A question therefore arises: how does this omission affect the
results of the nonadiabatic calculation of molecular properties?. In this
letter we will try to answer it.

Suppose that we try to approximate the energy of the system by minimization
of the variational energy $W=\left\langle \hat{H}_{T}\right\rangle
=\left\langle \varphi \right| \hat{H}_{T}\left| \varphi \right\rangle
/\left\langle \varphi \right| \left. \varphi \right\rangle $ as in the DEMO
method of Tachikawa et al\cite{TMNI98,TO00}. If $\varphi $ depends only on
translation--invariant coordinates then $W=W_{rel}=\left\langle \hat{H}%
\right\rangle $ because $\left\langle \hat{T}_{CM}\right\rangle =0$.
However, if $\varphi $ depends on the coordinates of the particles in the
laboratory--fixed set of axes, as in the case of the SCF wavefunction used
by Tachikawa et al (see, for example equations (10) and (7) in references
\cite{TMNI98} and \cite{TO00}, respectively), then $W=\left\langle \hat{T}%
_{CM}\right\rangle +\left\langle \hat{H}\right\rangle >W_{rel}$. From the
variational principle we know that $W_{rel}>E_{0}$, where $E_{0}$ is the
exact ground--state energy of the atomic or molecular system. Therefore, the
use of $\hat{H}_{T}$ (instead of $\hat{H}$) and a laboratory--fixed
set of axes for the electronic and nuclear coordinates in $\varphi $ will
result in an even larger estimation of the molecular energy.

It is well--known that the SCF wavefunction satisfies the virial theorem\cite
{FC87,TO00} $2\left\langle \hat{T}\right\rangle =-\left\langle
V\right\rangle $, but in this case we have a wrong relation because $%
\left\langle \hat{T}\right\rangle =\left\langle \hat{T}_{CM}\right\rangle
+\left\langle \hat{T}_{rel}\right\rangle >\left\langle \hat{T}%
_{rel}\right\rangle $. Therefore, under such conditions the virial theorem
may be a misleading indication of the quality of the wavefunction.

\begin{figure}[H]
\begin{center}
\includegraphics[width=9cm]{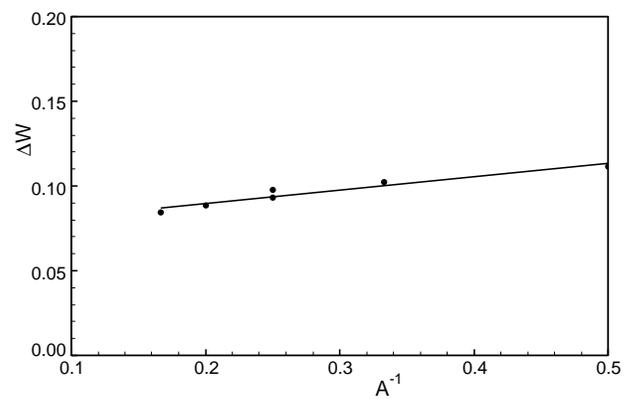}
\end{center}
\caption{$\Delta W$ vs. $A^{-1}$ for the H$_2$ isotopic series shown in
Table~\ref{tab:energies}.}
\label{fig:TCM}
\end{figure}

Table~\ref{tab:energies} shows the ground--state energies of some diatomic
molecules calculated with the internal Hamiltonian operator\cite{KW66,KA00}
and also the corresponding DEMO results of Tachikawa and Osamura\cite{TO00}
who did not remove the motion of the center of mass. As expected the
uncorrelated SCF energies are greater than those in which particle
correlation is explicitly taken into account\cite{KW66,KA00}. In addition to
it, we also expect the energy difference $\Delta W=W^{TO}-W^{KA}$ (where TO
and KA stand for Tachikawa and Osamura and Kinghorn and Adamowicz,
respectively) to depend on the expectation value $\left\langle \hat{T}%
_{CM}\right\rangle $ that should decrease as the molecular mass increases.
In fact, the third column of Table~\ref{tab:energies} shows this trend as
expected from the fact that $\left\langle \hat{T}_{CM}\right\rangle $ is
inversely proportional to the total molecular mass. If this argument were
correct then $\Delta W$ would exhibit an almost linear relation with the inverse
of the mass number
$A$. Fig.~\ref{fig:TCM} shows that this
is in fact the case for the values of the energy difference shown in Table~%
\ref{tab:energies}.

In order to illustrate (and in some way corroborate) the arguments above
we consider a simple but
nontrivial toy example given by the anharmonic oscillator
\begin{equation}
\hat{H}_{T}=-\frac{\hbar ^{2}}{2m_{1}}\frac{\partial ^{2}}{\partial x_{1}^{2}%
}-\frac{\hbar ^{2}}{2m_{2}}\frac{\partial ^{2}}{\partial x_{2}^{2}}%
+k(x_{1}-x_{2})^{4}  \label{eq:HT_osc}
\end{equation}
In terms of the relative $x=x_{1}-x_{2}$ and center--of--mass $%
X=(m_{1}x_{1}+m_{2}x_{2})/M$ coordinates, where $M=m_{1}+m_{2}$, we have
\begin{equation}
\hat{H}_{T}=-\frac{\hbar ^{2}}{2M}\frac{\partial ^{2}}{\partial X^{2}}-\frac{%
\hbar ^{2}}{2m}\frac{\partial ^{2}}{\partial x^{2}}+kx^{4}
\label{eq:HT_osc_CM}
\end{equation}
where $m=m_{1}m_{2}/M$ is the reduced mass. The first and second terms in
the right--hand--side of this equation are simple examples of the $\hat{T}%
_{CM}$ and $\hat{T}_{rel}$ operators, respectively, mentioned above. This
toy model may seem to be rather too unrealistic at first sight but if exhibits
some of the necessary features. First, it is separable into center of mass and
relative degrees of freedom. Second, we can apply simple variational
functions of coordinates defined in the laboratory--fixed set of axes as
well as functions of more convenient relative variables. Third, we can
calculate the eigenvalues of the relative Hamiltonian operator quite
accurately, which are useful for comparison.

To simplify the calculation we resort to the dimensionless coordinates $%
q_{i}=x_{i}/L $, where $L=[\hbar ^{2}/(m_{1}k)]^{1/6}$, and the total
dimensionless Hamiltonian operator
\begin{equation}
\hat{H}_{Td}=\frac{m_{1}L^{2}}{\hbar ^{2}}\hat{H}_{T}=-\frac{1}{2}\frac{%
\partial ^{2}}{\partial q_{1}^{2}}-\frac{\beta }{2}\frac{\partial ^{2}}{%
\partial q_{2}^{2}}+(q_{1}-q_{2})^{4}  \label{eq:HT_osc_d}
\end{equation}
where $\beta =m_{1}/m_{2}$. Analogously, the relative Hamiltonian operator
is given by
\begin{equation}
\hat{H}_{d}=-\frac{\beta +1}{2}\frac{\partial ^{2}}{\partial q^{2}}+q^{4}.
\label{eq:H_osc_d}
\end{equation}
where $q=q_1-q_2$ is the translation--invariant coordinate.

We first consider the variational function $\varphi _{r}(a,q)=\exp (-aq^{2})$%
, where $a$ is a variational parameter, and the total dimensionless
Hamiltonian operator (\ref{eq:HT_osc_d}). Notice that this trial function
depends only on the relative coordinate $q$. The
calculation is straightforward and we obtain $W_{r}=3\cdot 6^{1/3}(\beta
+1)^{2/3}/8$. Obviously, the optimized trial function satisfies the virial
theorem $\left\langle \hat{T}\right\rangle =\left\langle \hat{T}%
_{rel}\right\rangle =2\left\langle \hat{V}\right\rangle =6^{1/3}(\beta
+1)^{2/3}/4$.

In order to simulate an SCF function of the laboratory--fixed
coordinates we consider $\varphi _{nr}(a,b,q_{1},q_{2})=\exp
(-aq_{1}^{2}-bq_{2}^{2})$. The calculation is also straightforward and we
obtain $W_{nr}=3\cdot 6^{1/3}(\sqrt{\beta }+1)^{2}/[8(\sqrt{\beta }%
+1)^{2/3}]>W_{r}$. The optimized trial function also satisfies the virial
theorem $\left\langle \hat{T}\right\rangle =2\left\langle \hat{V}%
\right\rangle $, but in this case $\left\langle \hat{T}\right\rangle
>\left\langle \hat{T}_{rel}\right\rangle $ as discussed above.

\begin{figure}[H]
\begin{center}
\includegraphics[width=9cm]{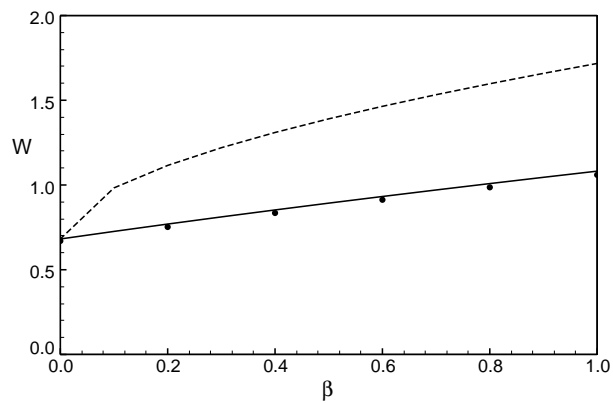}
\end{center}
\caption{Ground--state energy of the anharmonic oscillator calculated with
the variational function of the relative (solid line) and laboratory--fixed
(dashed line) coordinates and the accurate numerical results (circles).}
\label{fig:energies}
\end{figure}

Fig.~\ref{fig:energies} shows $W_{r}$, $W_{nr}$ and an accurate numerical
calculation of the ground--state energy of the dimensionless relative Hamiltonian
operator (\ref{eq:H_osc_d}) for $0<\beta <1$. We clearly appreciate the
advantage of using a trial wavefunction of internal coordinates, or of
properly removing the motion of the center of mass. We do not claim that the
error in the DEMO calculation of molecular energies\cite{TMNI98,TO00} is as
large as the one suggested by present anharmonic--oscillator, but this
simple model shows (at least) two aspects of the problem. First, that the energy
calculated by trial functions of the laboratory--fixed coordinates may be
considerably greater than those coming from the use of relative coordinates
if we do not remove the motion of the center of mass properly. And, second,
that the virial theorem is not a reliable indication of the quality of the
wavefunction if it is not based on the relative kinetic energy.

\begin{figure}[H]
\begin{center}
\includegraphics[width=9cm]{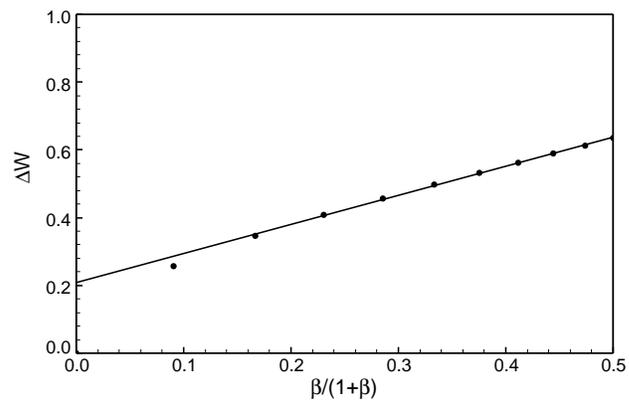}
\end{center}
\caption{$\Delta W$ vs. $\beta/(1+\beta)$ for the ground--state of the
anharmonic oscillator.}
\label{fig:energies2}
\end{figure}

We can carry out another numerical experiment with the toy model. The total
mass in units of $m_{1}$ is $M/m_{1}=(1+\beta )/\beta $. Fig.~\ref
{fig:energies2} shows that $\Delta W=W_{nr}-W_{r}$ depends almost linearly
on $\beta /(1+\beta )$ (at least for some values of $\beta $) as
suggested by the argument above about the actual molecular energies. We
appreciate that the toy model gives us another hint on the difference
between the
actual molecular energies calculated by Kinghorn and Adamowicz\cite{KA00}
and Tachikawa and Osamura\cite{TO00}.

Summarizing: if we do not properly separate the motion of the center of mass
in a calculation of atomic or molecular properties we expect inaccurate
results unless the approximate trial function depends only on internal,
translation--free coordinates. Otherwise, the effect of the kinetic energy
of the center of mass will be a too large estimate of the energy.
Under such conditions the virial theorem will result in a
misleading indication of a supposedly accurate wavefunction. These arguments
apply to the case in which all the particles are allowed to move\cite{TO00}
and may not be valid when some heavy particles\cite{TMNI98} (or all the
nuclei\cite{TO00}) are considered as merely point charges (a sort of clamped
nucleus approximation).

\begin{table}[H]
\caption{Nonadiabatic energies of some diatomic molecules}
\label{tab:energies}
\begin{center}
\begin{tabular}{lll}
\hline
Ref. & \multicolumn{1}{c}{$W$} & \multicolumn{1}{c}{$\Delta W$} \\ \hline
\multicolumn{3}{c}{H$_2$} \\ \hline
KA00 & -1.1640250232 & 0.111654 \\
TO00 & -1.052371 &  \\ \hline
\multicolumn{3}{c}{HD} \\ \hline
KW66 & -1.1654555 &  \\
KA00 & -1.1654718927 & 0.102116 \\
TO00 & -1.063356 &  \\ \hline
\multicolumn{3}{c}{HT} \\ \hline
KA00 & -1.1660020061 & 0.0987868 \\
TO00 & -1.068382 &  \\ \hline
\multicolumn{3}{c}{D$_2$} \\ \hline
KA00 & -1.1671688033 & 0.0918650 \\
TO00 & -1.074137 &  \\ \hline
\multicolumn{3}{c}{DT} \\ \hline
KA00 & -1.1678196334 & 0.0885406 \\
TO00 & -1.079279 &  \\ \hline
\multicolumn{3}{c}{T$_2$} \\ \hline
KA00 & -1.1685356688 & 0.0844127 \\
TO00 & -1.084123 &  \\ \hline
\end{tabular}
\end{center}
\end{table}

\end{document}